\documentclass[11pt]{article}

\usepackage{amsbsy,amsfonts,amssymb,amsmath}
\usepackage{graphics}
\usepackage{psfrag}
\usepackage[section]{placeins}

\newtheorem{theorem}{Theorem}[section]
\newtheorem{lemma}[theorem]{Lemma}

\newenvironment{proof}[1][Proof]{\begin{trivlist}
\item[\hskip \labelsep {\bfseries #1}]}{\end{trivlist}}
\newenvironment{definition}[1][Definition]{\begin{trivlist}
\item[\hskip \labelsep {\bfseries #1}]}{\end{trivlist}}

\newcommand{\qed}{\nobreak \ifvmode \relax \else
      \ifdim\lastskip<1.5em \hskip-\lastskip
      \hskip1.5em plus0em minus0.5em \fi \nobreak
      \vrule height0.75em width0.5em depth0.25em\fi}

\begin{document}

\title{An optimal bifactor approximation algorithm for
the metric uncapacitated facility location problem\footnote{
A short version of this paper appeared in the Proceedings of APPROX'07 \cite{Byrka07}.} }

\author{Jaroslaw Byrka\footnote{
 CWI, P.O. Box 94079, 1090 GB Amsterdam, The Netherlands,
 and Eindhoven University of Technology, P.O. Box 513, 5600 MB Eindhoven, The Netherlands
 (J.Byrka@cwi.nl, Karen.Aardal@cwi.nl).} \footnote{
Partially supported by the EU Marie Curie Research Training Network
ADONET, Contract No MRTN-CT-2003-504438.},
Karen Aardal$^\dagger$\footnote{Partially supported by the Dutch BSIK/BRICKS project.}
}

\maketitle

\begin{abstract}
We obtain a 1.5-approximation algorithm for the metric uncapacitated
facility location problem (UFL), which improves on the previously
best known 1.52-approximation algorithm by Mahdian, Ye and Zhang.
Note, that the approximability lower bound by Guha and Khuller is 1.463..

An algorithm is a {\em ($\lambda_f$,$\lambda_c$)-approximation
algorithm} if the solution it produces has total cost at most
$\lambda_f \cdot F^* + \lambda_c \cdot C^*$, where $F^*$ and $C^*$
are the facility and the connection cost of an optimal solution.
Our new algorithm, which is a modification of the
$(1+2/e)$-approximation algorithm of Chudak and Shmoys, is a
(1.6774,1.3738)-approximation algorithm for the UFL problem and is
the first one that touches the approximability limit curve
$(\gamma_f, 1+2e^{-\gamma_f})$ established by Jain, Mahdian
and Saberi. As a
consequence, we obtain the first optimal approximation algorithm
for instances dominated by connection costs. When combined with a
(1.11,1.7764)-approximation algorithm proposed by Jain et al.,
and later analyzed by Mahdian et al., we obtain
the overall approximation guarantee of 1.5 for the metric UFL
problem. We also describe how to use our algorithm to improve the
approximation ratio for the 3-level version of UFL.

\end{abstract}
{\bf Key words.} Facility location, approximation algorithms, LP-rounding.


\section{Introduction}
\subsection{Background on Uncapacitated Facility Location}
The Uncapacitated Facility Location (UFL) problem is defined as
follows. We are given a set $\mathcal{F}$ of \emph{facilities} and
a set $\mathcal{C}$ of \emph{clients}. For every facility $i \in
\mathcal{F}$, there is a nonnegative number $f_i$ denoting the
\emph{opening cost} of the facility. Furthermore, for every client
$j \in \mathcal{C}$ and facility $i \in \mathcal{F}$, there is a
\emph{connection cost} $c_{ij}$ between facility $i$ and client
$j$. The goal is to open a subset of the facilities $\mathcal{F}'
\subseteq \mathcal{F}$, and connect each client to an open
facility so that the total cost is minimized. The UFL problem is
NP-complete, and max SNP-hard (see~\cite{GuhaKhuller98}). A UFL
instance is \emph{metric} if its \emph{connection cost} function
satisfies the following variant of the \emph{triangle inequality}:
\begin{equation}\label{eq_triangleineq}
c_{ij} \leq c_{ij'} + c_{i'j'} + c_{i'j} \mbox{ for any } i,i' \in
\mathcal{C} \mbox{ and } j,j' \in \mathcal{F}\,.
\end{equation}

We will say that an algorithm is a $\lambda$-approximation
algorithm for a minimization problem if it computes, in polynomial
time, a solution that is at most $\lambda$ times more expensive
than the optimal solution. Specifically, for the UFL problem we
consider the notion of \emph{bifactor approximation} introduced by
Charikar and Guha~\cite{augmentation,augmentation2}.
We say that an algorithm is a
($\lambda_f$,$\lambda_c$)-approximation algorithm if the solution
it delivers has total cost at most $\lambda_f \cdot F^* +
\lambda_c \cdot C^*$, where $F^*$ and $C^*$ denote, respectively,
the facility and the connection cost of an optimal solution.
Note the potential ambiguity resulting from the possible existence of
multiple optimal solutions. When presenting our algorithm, we will
compare the solution cost only to the cost of the initial
fractional solution. Nevertheless, as we observe at the end of
Section~\ref{sec_alg}, adding an additional scaling step to our
algorithm is sufficient to get a guarantee in a comparison
with any feasible fractional solution.

Guha and Khuller~\cite{GuhaKhuller98} proved by a reduction from
Set Cover that there is no polynomial time $\lambda$-approximation
algorithm for the metric UFL problem with $\lambda < 1.463$,
unless $NP \subseteq DTIME(n^{\log \log n})$. Sviridenko showed
that the approximation lower bound of 1.463 holds, unless $P = NP$
(see~\cite{Vygen}).
Jain et al.~\cite{JMS} generalized the argument of Guha and
Khuller to show that the existence of a
($\lambda_f$,$\lambda_c$)-approximation algorithm with $\lambda_c
< 1+2e^{-\lambda_f}$ would imply $NP \subseteq DTIME(n^{\log \log
n})$.

The UFL problem has a rich history starting in the 1960's. The
first results on approximation algorithms are due to Cornu\'ejols,
Fisher, and Nemhauser~\cite{CFN77} who considered the problem with
an objective function of maximizing the ``profit'' of connecting
clients to facilities minus the cost of opening facilities. They
showed that a greedy algorithm gives an approximation ratio of
$(1-1/e)=0.632\dots$, where $e$ is the base of the natural
logarithm. This ratio was later improved to $0.828$
by Ageev and Sviridenko~\cite{AS99}.

For the objective function of minimizing the sum of
connection cost and opening cost, Hochbaum \cite{Hochbaum82}
presented a greedy algorithm with an $O(\log n)$ approximation
guarantee, where $n$ is the number of clients. By a
straightforward reduction from the Set Cover problem, it can be
shown that this cannot be improved unless $NP \subseteq
DTIME[n^{O(\log \log n)}]$ due to a result by Feige~\cite{feige}.
However, if the connection costs are restricted to satisfy the
triangle inequality (\ref{eq_triangleineq}), then constant
approximation guarantees can be obtained. In all results mentioned
below, except for the maximization objectives, it is assumed that
the costs satisfy these restrictions. If the distances between
facilities and clients are Euclidean, then for some location
problems approximation schemes have been obtained \cite{ARR98}.

The first approximation algorithm with constant approximation
ratio for the metric minimization problem
was developed by Shmoys, Tardos, and
Aardal~\cite{shmoys97approximation}. Since then numerous
improvements have been made. Guha and
Khuller~\cite{GuhaKhuller98,GK} introduced a \emph{greedy
augmentation procedure} (see also Charikar and Guha \cite{augmentation,augmentation2}).
A series of approximation algorithms
based on LP-rounding was then developed (see
e.g.~\cite{chudak98improved,ChudakS03,Sviridenko02}). There are
also greedy algorithms that only use the LP-relaxation implicitly
to obtain a lower bound for a primal-dual analysis. An example is
the JMS 1.61-approximation algorithm developed by Jain, Mahdian,
and Saberi \cite{JMS}. Some algorithms combine several techniques,
like the 1.52-approximation algorithm of Mahdian, Ye, and Zhang
\cite{mahdian-improved,MYZ}, which uses the JMS algorithm and the
greedy augmentation procedure. Up to now, their approximation
ratio of 1.52 was the best known.
Many more algorithms have been considered for the UFL problem and
its variants. We refer the interested reader to survey papers by
Shmoys~\cite{Shmoys00} and Vygen~\cite{Vygen}.

\subsection{Some basic techniques}
\label{sec_related_work}

In several LP-based approximation algorithms a \textit{clustering
step} is part of an algorithm for creating a feasible solution, see
Section \ref{sec_clustering} for more details. In this step a not
yet clustered client is chosen as the so-called ``cluster center''
and one of the facilities that fractionally serves the cluster
center in the LP solution is opened. Our main technique is to
modify the support graph corresponding to the LP solution before
clustering, and to use various average distances in the fractional
solution to bound the cost of the obtained solution.

A similar way of modifying
the LP-solution, called \textit{filtering}, was introduced by Lin and
Vitter~\cite{LV92}. Lin and Vitter
considered a broad class of 0-1 problems having both covering and packing
constraints.
They start by solving the LP-relaxation of the problem,
and in the subsequent filtering step
they select a subset of the variables that have positive value
in the LP solution and that have relatively large objective coefficients.
These variables are set equal to zero, which
results in a modified problem. The LP-relaxation of this modified problem
is then solved and rounding is applied.
In the paper by Shmoys et al.~\cite{shmoys97approximation}
filtering was also used in order to bound the connection costs. Here
again a subset of the variables that have a positive value in the
LP-solution are set equal to zero. The remaining positive variables
were scaled so as to remain feasible for the original LP-relaxation.

Later, Chudak \cite{chudak98improved} observed that the LP-relaxation
was already filtered in a certain sense as it is possible to state
that if a client is fractionally connected to a facility in the LP-solution,
then one can bound the cost of this connection in terms of the optimal LP-dual
variables. This observation was later used by Aardal, Chudak, and
Shmoys~\cite{Aardal99} in their algorithm for multi-level problems, and
by Sviridenko~\cite{Sviridenko02}. The filtering
done in our algorithm is slightly different
as the filtered LP-solution is not necessarily feasible with respect to the
LP-relaxation. Throughout this paper we will use the name
\emph{sparsening technique} for the combination of filtering with
our new analysis.

\subsection{Our contribution}\label{sec_contribution}
\begin{figure}

\includegraphics{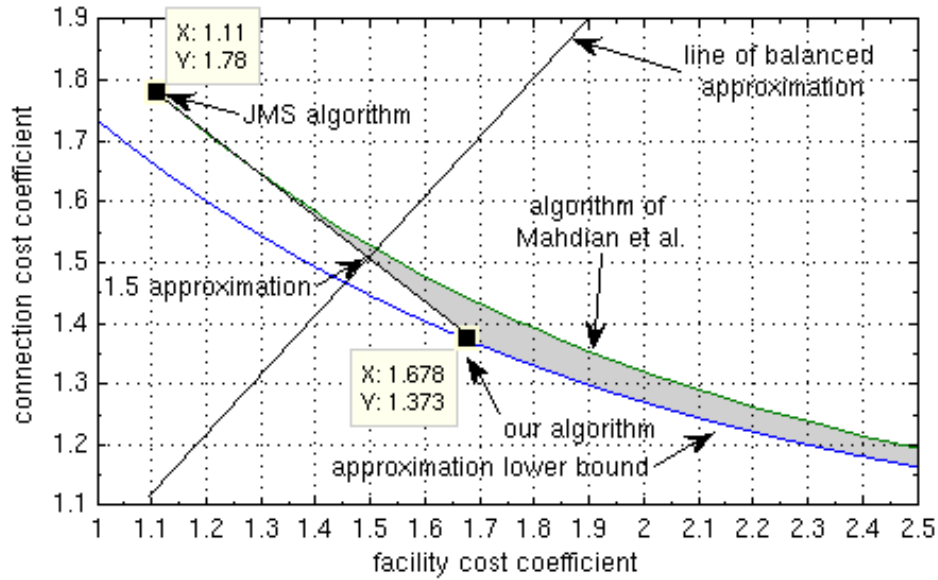}

\caption{Bifactor approximation picture. The gray area corresponds
to the improvement due to our algorithm.} \label{fig_contr}
\end{figure}

We modify the $(1+2/e)$-approximation algorithm of
Chudak~\cite{chudak98improved}, see also Chudak and
Shmoys~\cite{ChudakS03}, to obtain a new
(1.6774,1.3738)-approximation algorithm for the UFL problem. Our
linear programming (LP) rounding algorithm is the first one that
achieves an optimal bifactor approximation due to the matching
lower bound of $(\lambda_f, 1+2e^{-\lambda_f})$ established by
Jain et al.~\cite{JMS}. In fact we obtain an algorithm for each
point $(\lambda_f, 1+2e^{-\lambda_f})$ such that $\lambda_f \geq
1.6774$, which means that we have an optimal approximation
algorithm for instances dominated by connection cost (see
Figure~\ref{fig_contr}).

One of the main technical contributions of the paper is the proof
of Lemma~\ref{lem:main-lemma}, which gives a bound on the expected
connection cost in the case of using a path via cluster center
to connect a client. This lemma may potentially be useful in constructing
new algorithms for UFL and related problems.

One could view our contribution as an improved analysis of a minor
modification of the algorithm by Sviridenko~\cite{Sviridenko02},
which also introduces filtering to the algorithm of Chudak and
Shmoys. The filtering process that is used both in our algorithm
and in the algorithm by Sviridenko is relatively easy to describe,
but the analysis of the impact of this technique on the quality of
the obtained solution is quite involved in each case. Therefore,
we prefer to state our algorithm as an application of the
sparsening technique to the algorithm of Chudak and Shmoys, which
in our opinion is relatively easy do describe and analyze.

We start by observing that for a certain class of instances the
analysis of the algorithm of Chudak and Shmoys may be improved. We
call these instances \emph{regular}, and for the other instances
we propose a measure of their \emph{irregularity}. The goal of the
sparsening technique is to explore the irregularity of instances
that are potentially tight for the original algorithm of Chudak
and Shmoys. We cluster the given instance in the same way as in
the 1.58-approximation algorithm by
Sviridenko~\cite{Sviridenko02}, but we continue our algorithm in
the spirit of Chudak and Shmoys' algorithm, and we use certain
average distances to control the irregularities, which leads to an
improved bifactor approximation guarantee.

Our new algorithm may be combined with the
$(1.11,1.7764)$-approximation algorithm of Jain et al. to obtain a
1.5-approximation algorithm for the UFL problem. This is an
improvement over the previously best known 1.52-approxi\-mation
algorithm of Mahdian et al., and it cuts of a 1/3 off the gap with
the approximation lower bound by Guha and
Khuller~\cite{GuhaKhuller98}. An earlier version of this paper appeared in
\cite{Byrka07}.

We now give an informal sketch of our algorithm. Using this description
we give an outline of the paper.

\paragraph{Sketch of the algorithm.}
\begin{enumerate}
\item
    Solve the LP relaxation of the problem.
\item \label{point:mod}
    Modify the fractional solution by: 
\begin{itemize}
 \item scaling up the facility opening variables,
 \item modifying the connection variables to completely use the ``closest''
	fractionally open facilities,
 \item splitting facilities, if necessary, such that there is no slack
       between the amount that a client is assigned to a facility, and the
       amount by which this facility is opened.
\end{itemize}
\item \label{point:cluster}
    Divide clients into clusters based on the current fractional solution.
    In each cluster a specific client is assigned to be a ``cluster center''.
\item \label{point:open1}
    For every cluster, open one of the ``close'' facilities of the cluster center.
\item
    For each facility not considered above,
    open it independently with probability equal to the fractional opening.
\item
    Connect each client to an open facility that is closest to it.
\end{enumerate}

In Section \ref{sec_prelim} we give a brief overview of the main
ingredients of some known approximation algorithms for UFL. In
particular we state the LP relaxation of UFL, describe clustering,
scaling, and greedy augmentation. The clustering technique is common
for the existing LP-rounding algorithms for UFL,
and it is applied in Steps~\ref{point:cluster} and~\ref{point:open1} of the above algorithm.
Sparsening of the support graph of the LP solution, which is the essence
of Step~\ref{point:mod}, is discussed in Section~\ref{sec_sparsening},
where we also prove the crucial lemma on certain connection costs.
A more detailed description of the algorithm and its analysis are presented in Section~\ref{sec_alg},
and the 1.5-approximation algorithm is stated in Section~\ref{sec_approx}.
In Section~\ref{sec_multilevel} we show that the new
$(1.6774,1.3738)$-approximation algorithm may also be used to improve
the approximation ratio for the 3-level version of the UFL problem
to 2.492. A randomized approach to clustering is discussed
in Section~\ref{sec_randomized}, and, finally, in Section~\ref{sec_concluding} we present some
concluding remarks and open problems.

\section{Preliminaries}\label{sec_prelim}

We will review the concept of LP-rounding algorithms for the
metric UFL problem. These are algorithms that first solve the
linear relaxation of a given integer programming (IP) formulation
of the problem, and then round the fractional solution to produce
an integer solution with a value not too much higher than the
starting fractional solution. Since the optimal fractional
solution is at most as expensive as an optimal integral solution,
we obtain an estimation of the approximation factor.

\subsection{IP formulation and relaxation}

The UFL problem has a natural formulation as the following IP
problem.
\begin{eqnarray}
\nonumber
\mbox{min  }&\sum_{i \in \mathcal{F}, j \in \mathcal{C}}{c_{ij}x_{ij}} +
\sum_{i \in \mathcal{F}}{f_i y_i} &\\ \nonumber
&&\\ \nonumber
\mbox{s.t. } &\sum_{i \in \mathcal{F}} x_{ij} = 1 & \mbox{for all }j \in \mathcal{C}, \\ \nonumber
&x_{ij} - y_i \leq 0 & \mbox{for all } i \in \mathcal{F}, j \in \mathcal{C},\\ \label{eq_int}
&x_{ij},y_i \in \{ 0,1 \}       & \mbox{for all } i \in \mathcal{F}, j \in \mathcal{C}\,.
\end{eqnarray}

%
%

A linear relaxation of this IP formulation is obtained by
replacing the integrality constraints (\ref{eq_int}) by the
constraint
        $x_{ij} \geq 0 $     for all $i \in \mathcal{F}, j \in \mathcal{C}$ .
%
The value of the solution to this LP relaxation will serve as a
lower bound for the cost of the optimal solution. We will also
make use of the following dual formulation of this LP.

\begin{eqnarray*}
\mbox{max }& \sum_{ j \in \mathcal{C}}{v_j}&\\
&&\\[-.2cm]
\mbox{s.t. }    & \sum_{j \in \mathcal{C}} w_{ij} \leq f_i &\qquad \mbox{for all } i \in \mathcal{F},\\
        & v_j - w_{ij}  \leq c_{ij} &\qquad  \mbox{for all } i \in \mathcal{F}, j \in \mathcal{C},\\
    & w_{ij} \geq 0 &\qquad  \mbox{for all  } i \in \mathcal{F}, j \in \mathcal{C}\,.\\
\end{eqnarray*}

%
%

\subsection{Clustering}\label{sec_clustering}
The first constant factor approximation algorithm for the metric
UFL problem by Shmoys et al., but also the algorithms by Chudak
and Shmoys, and by Sviridenko are based on the following
clustering procedure.
Suppose we are given an optimal solution to the LP relaxation of
our problem. Consider the bipartite graph $G=((V',V''),E)$ with
vertices $V'$ being the facilities and $V''$ the clients of the
instance, and where there is an edge between a facility $i\in V'$
and a client $j\in V''$ if the corresponding variable $x_{ij}$ in
the optimal solution to the LP relaxation is positive. We call $G$
a \emph{support graph} of the LP solution. If two clients are both
adjacent to the same facility in graph $G$, we will say that they
are \emph{neighbors} in $G$.

\begin{figure}[t]
\begin{center}
\fboxsep7pt
\framebox[0.95\columnwidth]{
\begin{minipage}{0.90\columnwidth}
\begin{center}

\includegraphics{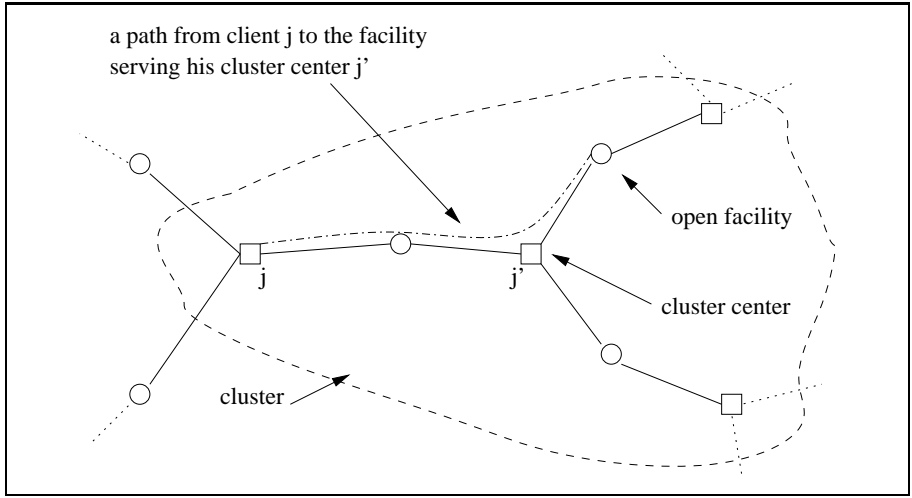}
\end{center}
\end{minipage}
}
\end{center}
\caption{A cluster. If we make sure
that at least one facility is open close to a cluster center $j'$ ,
then any other client $j$ from the cluster may use this facility.
Because the connection costs are assumed to be metric, the distance
to this facility is at most the length of the shortest path from $j$ to the
open facility.}
\label{fig:cluster}
\end{figure}

The clustering in this graph is a partitioning of clients into
clusters together with a choice of a leading client for each of
the clusters. This leading client is called a \emph{cluster
center}. Additionally we require that no two cluster centers are
neighbors in the support graph. This property helps us to open one
of the adjacent facilities for each cluster center.
For a picture of a cluster see Figure~\ref{fig:cluster}.

The algorithms by Shmoys et al., Chudak and Shmoys, and by
Sviridenko all use the following procedure to obtain the
clustering: While not all the clients are clustered, choose
greedily a new cluster center $j$, and build a cluster from $j$
and all the neighbors of $j$ that are not yet clustered. Obviously
the outcome of this procedure is a proper clustering. Moreover, it
has a desired property that clients are ``close'' to their cluster
centers. Each of the mentioned LP-rounding algorithms uses a
different greedy criterion for choosing new cluster centers. In
our algorithm we will use the clustering with the greedy criterion
of Sviridenko~\cite{Sviridenko02}. Another way of clustering is
presented in Section 7. 

\subsection{Scaling and greedy augmentation}\label{sub_greedy_aug}

The techniques described here are not directly used by our
algorithm, but they help to explain why the algorithm of Chudak
and Shmoys is close to optimal. We will discuss how scaling
facility opening costs before running an algorithm, together with
another technique, called \emph{greedy augmentation}, may help to
balance the analysis of an approximation algorithm for the UFL
problem.

The greedy augmentation technique introduced by Guha and
Khuller~\cite{GuhaKhuller98} (see also~\cite{augmentation,augmentation2}) is as
follows. Consider an instance of the metric UFL problem and a
feasible solution. For each facility $i \in \mathcal{F}$ that is
not opened in this solution, we may compute the amount of cost
that is saved by opening facility $i$, also called the \emph{gain}
of opening $i$, denoted by $g_i$. While there exists a facility
$i$ with positive gain $g_i$, the greedy augmentation procedure
opens a facility that maximizes the ratio of gain to the facility
opening cost $\frac{g_i}{f_i}$, and updates the remaining values
of $g_i$.

Suppose we are given an approximation algorithm $A$ for the metric
UFL problem and a real number $\delta \geq 1$. Consider the
following algorithm $S_\delta(A)$.
\begin{enumerate}
\item scale up all facility opening costs by a factor $\delta$;
\item run algorithm $A$ on the modified instance;
\item scale back the opening costs;
\item run the greedy augmentation procedure.
\end{enumerate}

Following the analysis of Mahdian, Ye, and
Zhang~\cite{mahdian-improved} one may prove the following lemma.
\begin{lemma} \label{lem:scaling}
Suppose $A$ is a ($\lambda_f$,$\lambda_c$)-approximation algorithm
for the metric UFL problem, then $S_\delta(A)$ is a ($\lambda_f +
ln(\delta)$,$1+ \frac{\lambda_c -1}{\delta}$)-approximation
algorithm for this problem.
\end{lemma}

This method may be applied to balance an
($\lambda_f$,$\lambda_c$)-approximation algorithm with $\lambda_f
<< \lambda_c$. However, our 1.5-approximation algorithm is
balanced differently. It is a composition of two algorithms that
have opposite imbalances.

\section{Sparsening the graph of the fractional solution}\label{sec_sparsening}
In this section we describe a technique that we use to control the
expected connection cost of the obtained integer solution. Our
technique is based on the concept of \emph{filtering}, introduced
by Lin and Vitter~\cite{LV92}, see Section \ref{sec_related_work}.
We will give an alternative
analysis of the effect of filtering on a fractional solution to
the LP relaxation of the UFL problem.

Suppose that, for a given UFL instance, we have solved its LP
relaxation, and that the optimal primal solution is $(x^*,y^*)$
and the corresponding optimal dual solution is $(v^*,w^*)$. Such a
fractional solution has facility cost $F^* = \sum_{i \in
\mathcal{F}} {f_i y^*_i}$ and connection cost $C^* = \sum_{i \in
\mathcal{F}, j \in \mathcal{C}}{c_{ij}x^*_{ij}}$. Each client $j$
has its share $v^*_j$ of the total cost. This cost may again be
divided into a client's fractional connection cost $C^*_j = \sum_{
i \in \mathcal{F}} {c_{ij}x^*_{ij}}$, and its fractional facility
cost $F^*_j = v^*_j - C^*_j$.

\subsection{Motivation and intuition}

The idea behind the sparsening technique is to make use of
irregularities of an instance if they occur. We call an instance
\emph{locally regular} around client $j$ if the facilities that
serve $j$ in the fractional solution $(x^*,y^*)$ are all at the
same distance from $j$. An instance which is locally regular
around each client is called \emph{regular}. We begin by observing
that for such an instance the algorithm of Chudak and Shmoys
produces a solution whose cost is bounded by $F^* +
(1+\frac{2}{e})C^*$, which is an easy consequence of the original
analysis~\cite{ChudakS03}, but also follows from our analysis in
Section~\ref{sec_alg}. Although this observation might not be very
powerful itself, the value $F^* + (1+\frac{2}{e})C^*$ happens to
be the intersection point between the bifactor approximation lower
bound curve $(\lambda_f, 1+2e^{-\lambda_f})$ and the $y$-axis in
Figure \ref{fig_contr}. Moreover, for regular instances we may
apply the technique described in Section~\ref{sub_greedy_aug} to
obtain an approximation algorithm corresponding to any single
point on this curve. In particular, we may simply use this
construction to get an optimal $1.463\ldots$-approximation
algorithm for regular instances of the metric UFL problem. Note,
that the proof of the matching hardness of approximation also uses
instances that are essentially\footnote{ These instances come from
a reduction from the SET COVER problem. Clients represent elements
to be covered, and facilities represent subsets. The distance
$c_{ij}$ equals 1 if subset $i$ contains element $j$ and it equals
$3$ otherwise. To formally argue about the regularity of such an
instance we would need to construct an optimal fractional solution
using only facilities at distance 1. } regular.

The instances that are not regular are called \emph{irregular} and
these are the instances for which it is more difficult to create a
feasible integer solution with good bounds on the connection cost.
In fractional solutions of irregular instances there exist clients
that are fractionally served by facilities at different distances.
Our approach is to divide facilities serving a client into two
groups, namely \emph{close} and \emph{distant} facilities. We will
remove links to distant facilities before the clustering step, so
that if there are irregularities, then distances to cluster
centers will decrease.

We measure the local irregularity of an instance by comparing the
fractional connection cost of a client to the average distance to
its  distant facilities. In the case of a regular instance, the
sparsening technique gives the same results as the technique
described in section~\ref{sub_greedy_aug}, but for irregular
instances sparsening makes it possible to construct
an integer solution with a better bound on the connection costs.

\subsection{Details} \label{subsection_details}

We will start by modifying the optimal fractional LP-solution
$(x^*,y^*)$ by scaling the $y$-variables by a constant $\gamma >
1$ to obtain a fractional solution $(x^*,\tilde{y} )$, where $
\tilde{y} = \gamma \cdot y^*$.
Note that by scaling we might set some $\tilde{y}_i > 1$.
In the filtering of Shmoys et al.~such a variable
would instantly be rounded to 1. However, for the compactness
of a later part of our analysis it is important not to round
these variables, but rather to split facilities. Before we discuss
splitting, let us fist modify the connection variables.
A version of this argument, which describes all these modifications of the fractional solution
at once, is given in~\cite{Sviridenko02}[Lemma 1].

Suppose that the values of the
$y$-variables are scaled and fixed, but that we now have the freedom to
change the values of the $x$-variables in order to minimize the
connection cost.
For each client $j$ we compute the values of the corresponding
$\tilde{x}$-variables in the following way. We choose an ordering
of facilities with nondecreasing distances to client $j$. We
connect client $j$ to the first facilities in the ordering so that
among the facilities fractionally serving $j$, only the last one
in the chosen ordering may be opened by more than that it serves
$j$. Formally, for any facilities $i$ and $i'$ such that $i'$ is
later in the ordering, if $\tilde{x}_{ij} < \tilde{y}_i$ then
$\tilde{x}_{i'j}=0$.

In the next step, we eliminate the occurrences of
situations where $0 < \tilde{x}_{ij} < \tilde{y}_i$. We do so by
creating an equivalent instance of the UFL problem, where facility
$i$ is split into two identical facilities $i'$ and $i''$. In the
new setting, the opening of facility $i'$ is $\tilde{x}_{ij}$ and
the opening of facility $i''$ is $\tilde{y}_i - \tilde{x}_{ij}$.
The values of the $\tilde{x}$-variables are updated accordingly.
By repeatedly applying this procedure we obtain a so-called
\emph{complete} solution $(\overline{x}, \overline{y})$, i.e., a
solution in which no pair $i \in \mathcal{F}, j \in \mathcal{C}$
exists such that $0 < \overline{x}_{ij} < \overline{y}_i$
(see~\cite{Sviridenko02}[Lemma 1] for a more detailed argument).

In the new complete solution $(\overline{x}, \overline{y})$ we
distinguish groups of facilities that are especially important for
a particular client. For a client $j$ we say that a facility $i$
is one of its \emph{close facilities} if it fractionally serves
client $j$ in $(\overline{x}, \overline{y})$; $\mathcal{C}_j=\{ i
\in \mathcal{F} | \overline{x}_{ij}>0 \}$ is the set of close
facilities of $j$. If $\overline{x}_{ij}=0$, but facility $i$ was
serving client $j$ in solution $(x^*,y^*)$, then we say, that $i$
is a \emph{distant} facility of client $j$; $\mathcal{D}_j=\{ i
\in \mathcal{F} | \overline{x}_{ij}=0, x^*_{ij}>0 \}$ is the set
of distant facilities of $j$.

We will extensively use the average distances between single clients and groups of facilities
defined as follows.
\begin{definition}
  For any client $j \in \mathcal{C}$, and for any subset of facilities $\mathcal{F'} \subset \mathcal{F}$
  such that $\sum_{i \in \mathcal{F'}}{y^*_i} > 0$, let
\[
    d(j,\mathcal{F'}) = \frac{\sum_{i \in \mathcal{F'}}{c_{ij} \cdot y^*_i}}
                           {\sum_{i \in \mathcal{F'}}{y^*_i}}.
\]
\end{definition}

To interpret differences between certain average distances we will
use the following parameter.

\begin{definition}Let
\[
r_\gamma(j) = \left\{
 \begin{array}{cl}
 \frac{d(j,\mathcal{D}_j) - d(j,\mathcal{D}_j \cup \mathcal{C}_j)}{F^*_j} &
  \mbox{  for } F^*_j > 0 \\
  0 & \mbox{  for } F^*_j = 0 .\\
\end{array}\right.
\]
\end{definition}

The value $r_\gamma(j)$ is a measure of the irregularity of the
instance around client $j$. It is the average distance to a
distant facility minus the fractional connection cost $C^*_j$
(note, that $C^*_j = d(j,\mathcal{D}_j \cup \mathcal{C}_j)$ is the
general average distance to both close and distant facilities)
divided by the fractional facility cost of a client $j$; or it is
equal to $0$ if $F^*_j = 0$. Since $d(j, \mathcal{D}_j) \leq
v_j^*$, $C^*_j = d(j,\mathcal{D}_j \cup \mathcal{C}_j)$ and $C_j^*
+ F_j^* = v_j^*$, $r_\gamma(j)$ takes values between $0$ and $1$.
$r_\gamma(j)=0$ means that client $j$ is served in the solution
$(x^*,y^*)$ by facilities that are all at the same distance. If
$r_\gamma(j)=1$, then the facilities are at different distances
and the distant facilities are all so far from $j$ that $j$ is not
willing to contribute to their opening. In fact, for clients $j$
with $F^*_j = 0$ the value of $r_\gamma(j)$ is not relevant for
our analysis.


Consider yet another quantity, namely $r_\gamma'(j) = r_\gamma(j)
* (\gamma - 1)$. Observe, that for a client $j$ with $F^*_j > 0$
we have
\begin{displaymath}r_\gamma'(j) = \frac{ d(j,\mathcal{D}_j
\cup \mathcal{C}_j) - d(j,\mathcal{C}_j) }{F^*_j}\,.
\end{displaymath}

\begin{figure}[t]
\begin{center}
\fboxsep7pt
\framebox[0.95\columnwidth]{
\begin{minipage}{0.90\columnwidth}
\begin{center}

\psfrag{g}{$\frac{1}{\gamma}$}
\psfrag{c1}{$v^*_j = C^*_j + F^*_j$}
\psfrag{c2}{$C^*_j + r_\gamma(j) \cdot F^*_j$}
\psfrag{c3}{$C^*_j$}
\psfrag{c4}{$C^*_j - r'_\gamma(j) \cdot F^*_j$}

\includegraphics{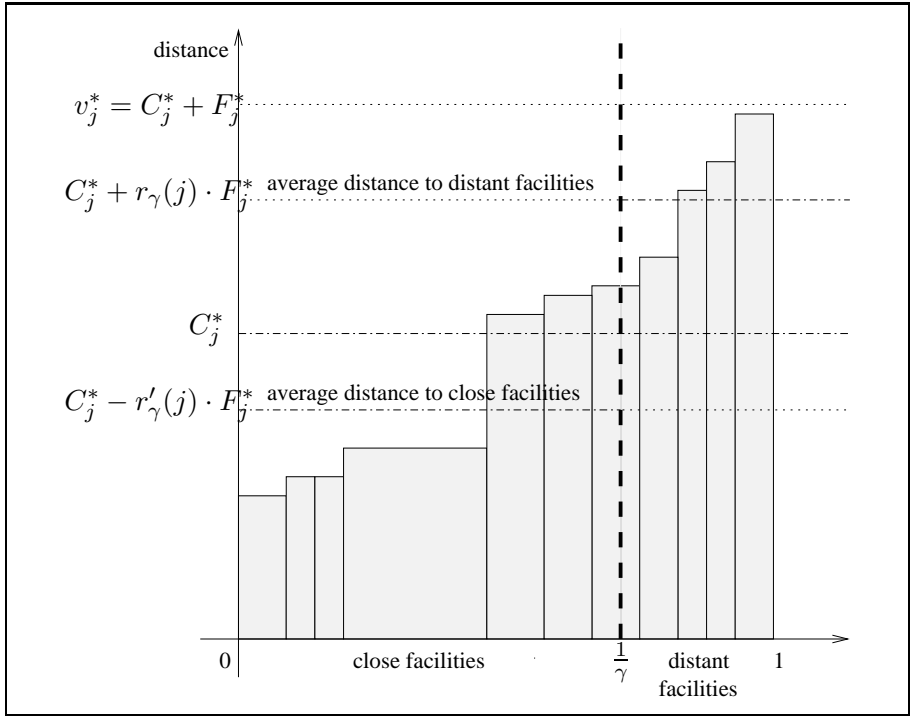}
\end{center}
\end{minipage}
}
\end{center}
\caption{Distances to facilities serving client $j$; the width of
a rectangle corresponding to facility $i$ is equal to $x^*_{ij}$.
The figure explains the meaning of  $r_\gamma(j)$ and
$r_\gamma'(j)$.} \label{fig:distances}
\end{figure}

We may use the definitions of $r_\gamma(j)$ and $r_\gamma'(j)$ together with
$C_j^*=d(j,\mathcal{D}_j\cup\mathcal{C}_c)$ to rewrite some distances from client $j$
in the following form (see also Figure~\ref{fig:distances}):
\begin{itemize}
\item
    the average distance to a close facility is
    \[
    D_{av}^C(j) = d(j, \mathcal{C}_j) = C^*_j - r_\gamma'(j) \cdot F^*_j,
    \]
\item
  the average distance to a distant facility is
  \[
  D_{av}^D(j) = d(j, \mathcal{D}_j) = C^*_j + r_\gamma(j) \cdot F^*_j,
  \]
\item
    the maximal distance to a close facility is
    \[
    D_{max}^C(j) \leq D_{av}^D(j) = C^*_j + r_\gamma(j) \cdot F^*_j.
    \]
\end{itemize}

In the following lemma we will prove an upper bound on the average
distance from client $j$ to another group of facilities.

\begin{lemma} \label{lem:main-lemma}
Suppose $\gamma < 2$ and that clients $j,j' \in \mathcal{C}$ are
neighbors in $(\overline{x},\overline{y})$, i.e. $\exists i \in
\mathcal{F} \mbox{ s.t. } \overline{x}_{ij} > 0 \mbox{ and }
\overline{x}_{ij'} > 0 $. Then, either $\mathcal{C}_{j'} \setminus
(\mathcal{C}_j \cup \mathcal{D}_j ) = \emptyset $ or
\[
 d(j,\mathcal{C}_{j'} \setminus (\mathcal{C}_j \cup \mathcal{D}_j )) \leq
 D_{av}^D(j) + D_{max}^C(j') + D_{av}^C(j').
\]
\end{lemma}

\begin{proof}
Assume that $\mathcal{C}_{j'} \setminus (\mathcal{C}_j \cup
\mathcal{D}_j )$ is not empty, since otherwise we are done.

\paragraph{Case 1.} Assume that the distance between $j$ and $j'$ is at most $D_{av}^D(j) + D_{av}^C(j')$.
By a simple observation, that a maximum is larger that the
average, we get
\begin{equation}\label{eq:c1}
 d(j',\mathcal{C}_{j'} \setminus (\mathcal{C}_j \cup \mathcal{D}_j )) \leq D_{max}^C(j').
\end{equation}
Combining the assumption with~(\ref{eq:c1}), we obtain
\begin{displaymath}
d(j,\mathcal{C}_{j'} \setminus (\mathcal{C}_j \cup \mathcal{D}_j
)) \leq
 D_{av}^D(j) + D_{max}^C(j') + D_{av}^C(j')\,.
\end{displaymath}

\paragraph{Case 2.} Assume that the distance between $j$ and $j'$ is longer than $D_{av}^D(j) + D_{av}^C(j')$.
Since $d(j, \mathcal{C}_j \cap \mathcal{C}_{j'}) \leq D_{av}^D(j)$, the assumption implies
\begin{equation} \label{eq:1}
d(j', \mathcal{C}_j \cap \mathcal{C}_{j'}) > D_{av}^C(j').
\end{equation}
Consider the following two sub-cases.

\paragraph{Case 2a.} Assume that $d(j',\mathcal{C}_{j'} \cap \mathcal{D}_j) \geq D_{av}^C(j')$.\\
This assumption together with~(\ref{eq:1}) gives
\begin{equation} \label{eq:2'}
d(j', \mathcal{C}_{j'} \cap(\mathcal{C}_j \cup \mathcal{D}_j)) \geq D_{av}^C(j').
\end{equation}
Recall that $D_{av}^C(j') = d(j', \mathcal{C}_{j'})$.
Hence~(\ref{eq:2'}) is equivalent to
\begin{equation} \label{eq:2}
d(j', \mathcal{C}_{j'} \setminus  (\mathcal{C}_j \cup \mathcal{D}_j)) \leq D_{av}^C(j').
\end{equation}
Since $j$ and $j'$ are neighbors, the distance between them is at
most $D_{max}^C(j) + D_{max}^C(j')$. By the triangle inequality
(\ref{eq_triangleineq}) we may add this distance to ~(\ref{eq:2})
and get \begin{displaymath}
d(j,\mathcal{C}_{j'} \setminus
(\mathcal{C}_j \cup \mathcal{D}_j )) \leq
 D_{av}^D(j) + D_{max}^C(j') + D_{av}^C(j')\,.
 \end{displaymath}

\paragraph{Case 2b.} In the remaining case we assume that
$d(j',\mathcal{C}_{j'} \cap \mathcal{D}_j) < D_{av}^C(j')$.\\
This assumption may also be written as
\begin{equation} \label{eq:3}
d(j',\mathcal{C}_{j'} \cap \mathcal{D}_j) = D_{av}^C(j') - z \mbox{ for some }z > 0.
\end{equation}
Now we combine~(\ref{eq:3}) with the assumption of Case 2 to get
\begin{equation} \label{eq:4}
d(j,\mathcal{C}_{j'} \cap \mathcal{D}_j) \geq D_{av}^D(j) + z.
\end{equation}
Let $\hat{y} = \sum_{i\in (\mathcal{C}_{j'} \cap \mathcal{D}_j)}{\overline{y}_i}$
be the total fractional opening of facilities in
$\mathcal{C}_{j'} \cap \mathcal{D}_j$ in the modified fractional
solution $(\overline{x}, \overline{y})$.\\

Observe that~(\ref{eq:4}) together with the definition
$d(j,\mathcal{D}_j) = D_{av}^D(j)$ implies that the set
$(\mathcal{D}_j \setminus \mathcal{C}_{j'})$ is not empty.
Moreover it contains facilities whose opening variables
$\overline{y}$ sum up to $\gamma -1 -\hat{y} > 0$. More precisely,
inequality~(\ref{eq:4}) implies $d(j, \mathcal{D}_j \setminus
\mathcal{C}_{j'}) \leq D_{av}^D(j) - z \cdot \frac{\hat{y}}{\gamma
-1 -\hat{y}}$. Hence
\begin{equation} \label{eq:5}
D_{max}^C(j) \leq D_{av}^D(j)- z \cdot \frac{\hat{y}}{\gamma -1 -\hat{y}}.
\end{equation}
We combine~(\ref{eq:5}) with the assumption of Case 2 to conclude
that the minimal distance from $j'$ to a facility in
$\mathcal{C}_{j'} \cap \mathcal{C}_j$ is at least $D_{av}^D(j) +
D_{av}^C(j') - D_{max}^C(j) \geq D_{av}^C(j') + z \cdot
\frac{\hat{y}}{\gamma -1 -\hat{y}}$. Hence
\begin{equation} \label{eq:6}
d(j',\mathcal{C}_{j'} \cap \mathcal{C}_j) \geq D_{av}^C(j') + z \cdot \frac{\hat{y}}{\gamma -1 -\hat{y}}.
\end{equation}
Recall that, by definition, $d(j', \mathcal{C}_{j'}) D_{av}^C(j')$. Hence equality~(\ref{eq:3}) may be written as
\begin{equation} \label{eq:7'}
d(j', \mathcal{C}_{j'} \setminus \mathcal{D}_j) = D_{av}^C(j') + z \cdot \frac{\hat{y}}{1-\hat{y}}.
\end{equation}
Since, by the assumption that $\gamma < 2$, we have
$\frac{\hat{y}}{1-\hat{y}} < \frac{\hat{y}}{\gamma -1 -\hat{y}}$,
we may also write
\begin{equation} \label{eq:7}
d(j', \mathcal{C}_{j'} \setminus \mathcal{D}_j) < D_{av}^C(j') + z \cdot \frac{\hat{y}}{\gamma -1 -\hat{y}}.
\end{equation}
We may now combine~(\ref{eq:7}) with~(\ref{eq:6}) to get
\begin{equation} \label{eq:8}
d(j', \mathcal{C}_{j'} \setminus (\mathcal{D}_j \cup \mathcal{C}_j)) < D_{av}^C(j') + z \cdot \frac{\hat{y}}{\gamma -1 -\hat{y}}.
\end{equation}
Finally, we bound the distance form $j$ to $j'$ by $D_{max}^C(j) + D_{max}^C(j')$ to get
\[
\begin{array}{rl}
d(j,\mathcal{C}_{j'} \setminus (\mathcal{C}_j \cup \mathcal{D}_j ))
& \leq D_{max}^C(j) + D_{max}^C(j') + d(j', \mathcal{C}_{j'} \setminus (\mathcal{D}_j \cup \mathcal{C}_j))  \\
& \leq D_{av}^D(j)- z \cdot \frac{\hat{y}}{\gamma -1 -\hat{y}} + D_{max}^C(j') + D_{av}^C(j') + z \cdot \frac{\hat{y}}{\gamma -1 -\hat{y}} \\
& = D_{av}^D(j) + D_{max}^C(j') + D_{av}^D(j')\,,
\end{array}
\]
where the second inequality is an application of~(\ref{eq:8}) and~(\ref{eq:5}).
\qed
\end{proof}


\section{Our new algorithm}\label{sec_alg}

Here we again state our algorithm (cf.~Section \ref{sec_contribution}), but now
we use the notation developed in the previous sections.

\vspace{.3cm}
\noindent\textbf{Algorithm $A1(\gamma)$}:
\begin{enumerate}
\item
    Solve the LP relaxation of the problem to obtain a solution $(x^*,y^*)$.
\item
    Modify the fractional solution as described in Section~\ref{subsection_details}
    to obtain a complete solution $(\overline{x}, \overline{y})$.
\item
  Compute a greedy clustering for the solution $(\overline{x}, \overline{y})$,
  choosing as cluster centers unclustered clients minimizing $D_{av}^C(j) + D_{max}^C(j)$.
\item \label{corelating}
    For every cluster center $j$, open one of its close facilities randomly
    with probabilities $\overline{x}_{ij}$.
\item
    For each facility $i$ that is not a close facility of any cluster center,
    open it independently with probability $ \overline{y}_i$.
\item
    Connect each client to an open facility that is closest to it.
\end{enumerate}

Consider the binary vector $y \in \{0,1\}^{|\mathcal{F}|}$ encoding the facilities
opened in Steps 4 and 5 of Algorithm $A1(\gamma)$.
With the following lemma we give an upper bound on the expected distance from
a client to the
closest of the facilities opened by the algorithm within a certain subset of facilities.

\begin{lemma} \label{exp_distance_lemma}
 Given are a random vector $y \in \{0,1\}^{|\mathcal{F}|}$ produced by Algorithm
 $A1(\gamma)$, a subset $A\subseteq \mathcal{F}$ of facilities such that $\sum_{i\in A}\bar{y}_i>0$,
 and a client $j\in\mathcal{C}$. Then, the following holds:
\[
 E\left[\min_{i\in A, y_i=1}c_{ij} \ | \ \sum_{i\in A} y_i \geq 1\right] \leq d(j,A)
\]
\end{lemma}
\begin{proof}
  Observe, that the opening of facilities from $A$ is either pairwise independent,
  or there exist disjoint subsets $A_1, A_2, \ldots \subseteq A$, which correspond to
  clusters created in Step 3 of the algorithm, such that the opening of facilities
  in each $A_k$ is negatively correlated but facilities from different sets are uncorrelated.
  The correlation in these subsets is a result of Step~\ref{corelating} of the algorithm.
  In each such $A_k$, there is at most 1 facility opened, and the probability that one is
  opened equals $\sum_{i \in A_k}\overline{y}_i$. Therefore, for the purpose of this proof, we may replace
  each $A_k$ by a new facility $i_k$ with distance to $j$ equal $d(j,A_k)$ and fractional opening
  $\overline{y}_{i_k} = \sum_{i \in A_k}\overline{y}_i$.
  After this replacement for each $A_k$, we have a set of facilities that are opened independently.

  Consider the facilities from $A$ in the order $i_1, i_2, \ldots$ of nondecreasing distance from $j$.
  Since their opening is independent, the probability that $i_l$
  counts as closest among the open facilities is
\begin{eqnarray*}
  p_l  &=& Pr[y_{i_1}=0]\cdot Pr[y_{i_2}=0]\cdot \ldots \cdot Pr[y_{i_{(l-1)}}=0] \cdot Pr[y_{i_l}=1] \\
 &=& (1-\overline{y}_{i_1})(1-\overline{y}_{i_2})\cdot \ldots \cdot (1-\overline{y}_{i_{(l-1)}}) \cdot \overline{y}_{i_l}.
\end{eqnarray*}
The expected distance may be bounded as:
\begin{eqnarray*}
 E\left[\min_{i\in A, y_i=1}c_{ij} \ | \ \sum_{i\in A} y_i \geq 1\right]
& = & \sum_{l=1}^{|A|} p_l c_{i_lj} \\
& = & \sum_{l=1}^{|A|} \left(\Pi_{o=1}^{l-1}(1-\overline{y}_{i_o})\right) \overline{y}_{i_l} c_{i_l,j} \\
& = & \frac{\sum_{l=1}^{|A|} \left(\Pi_{o=1}^{l-1}(1-\overline{y}_{i_o})\right) \overline{y}_{i_l} c_{i_l,j} }
		{\sum_{l=1}^{|A|} \left(\Pi_{o=1}^{l-1}(1-\overline{y}_{i_o})\right) \overline{y}_{i_l}}\\
& \leq & \frac{\sum_{l=1}^{|A|} \overline{y}_{i_l} c_{i_l,j} }
		{\sum_{l=1}^{|A|} \overline{y}_{i_l}} \\
& = &  \frac{\sum_{i \in A} \overline{y}_{i_l} c_{i_l,j} }
	{\sum_{i \in A} \overline{y}_{i_l}} = d(j,A).
\end{eqnarray*}
The second equality comes from the fact that,
under the condition that $\sum_{i\in A} y_i > 1$, the sum of probabilities $\sum p_l$ equals 1.
The inequality is a comparison of weighted arithmetical averages, where the first
one has lower weights for bigger elements.
\qed
\end{proof}

In the analysis of our algorithm we will also use the following result:
\begin{lemma} \label{probability_lemma}
    Given are $n$ independent events that occur with probabilities $p_1, p_2, \ldots, p_n$ respectively.
    The probability that at least one of these events occurs is at least equal to
    $1 - \frac{1}{e^{\sum_{i=1}^n p_i}}$, where $e$ denotes the base of the natural logarithm.
\end{lemma}

Let $\gamma_0$ be defined as the only positive solution to the following equation.
\begin{equation} \label{eq:gamma0}
\frac{1}{e} + \frac{1}{e^{\gamma_0}} - (\gamma_0 - 1) \cdot (1 - \frac{1}{e} + \frac{1}{e^{\gamma_0}}) = 0
\end{equation}
An approximate value of this constant is $\gamma_0 \approx 1.67736$.
As we will observe in the proof of Theorem~\ref{thm:main-theorem},
equation~(\ref{eq:gamma0}) appears naturally in the analysis of algorithm $A1(\gamma)$.

\begin{theorem} \label{thm:main-theorem}
  Algorithm $A1(\gamma_0)$ produces a solution with expected cost\\
 $E[cost(SOL)] \leq \gamma_0 \cdot F^* + 1 + \frac{2}{e^{\gamma_0}} \cdot C^*$.
\end{theorem}

\begin{proof}
The expected facility opening cost of the solution is \\
$E[F_{SOL}] = \sum_{i \in \mathcal{F}}{f_i \overline{y}_i} \gamma \cdot \sum_{i \in \mathcal{F}}{f_i y_i^*} = \gamma \cdot F^*$.

To bound the expected connection cost we show that for each client $j$ there is an open facility
within a certain distance with a certain probability.
If $j$ is a cluster center, one of its close facilities is open and the expected distance
to this open facility is $D_{Ave}^C(j) = C^*_j - r_\gamma'(j) \cdot F^*_j \leq C^*_j$.

\begin{figure}[t]
\begin{center}
\fboxsep7pt
\framebox[0.95\columnwidth]{
\begin{minipage}{0.90\columnwidth}
\begin{center}

\includegraphics{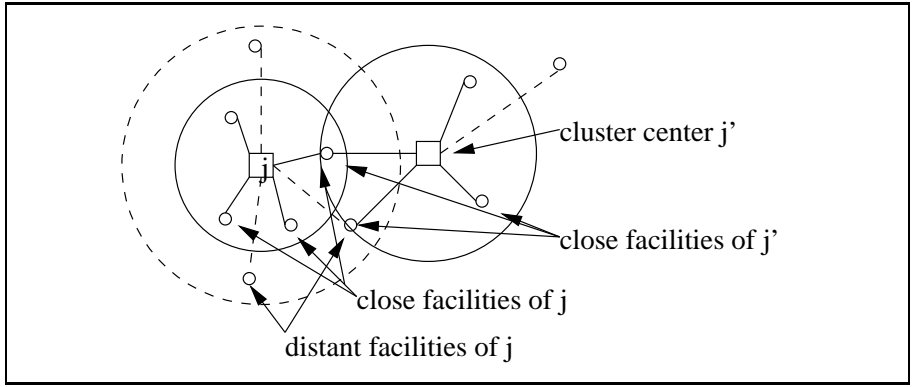}
\end{center}
\end{minipage}
}
\end{center}
\caption{Facilities that client $j$ may consider: its close facilities, distant facilities, and close
facilities of cluster center $j'$.}
\label{fig:routing}
\end{figure}

If $j$ is not a cluster center, it first considers its close facilities (see Figure~\ref{fig:routing}).
If any of them is open, by Lemma~\ref{exp_distance_lemma} the expected distance to the closest open facility is at most $D_{av}^C(j)$.
From Lemma~\ref{probability_lemma}, at least one close facility is open with probability $p_c \geq (1-\frac{1}{e})$.

Suppose none of the close facilities of $j$ is open, but at least one of its distant facilities is open.
Let $p_d$ denote the probability of this event. Again by Lemma~\ref{exp_distance_lemma}, the expected distance to the closest facility is then at most $D_{av}^D(j)$.

If neither any close nor any distant facility of client $j$ is open,
then $j$ may connect itself to the facility serving its cluster center $j'$.
Again from Lemma~\ref{probability_lemma}, such an event happens with probability $p_s \leq \frac{1}{e^\gamma}$.
We will now use the fact that if $\gamma < 2$ then, by Lemma~\ref{lem:main-lemma} and Lemma~\ref{exp_distance_lemma},
the expected distance from $j$ to the facility opened around $j'$ is at most
$D_{av}^D(j) + D_{max}^C(j') + D_{av}^C(j')$.

Finally, we combine the probabilities of particular cases
with the bounds on the expected connection for each of the cases
to obtain the following upper bound on the expected total connection cost.

\begin{center}
\begin{tabular}{ r l }

$E[C_{SOL}]$
     & $\leq \sum_{j \in \mathcal{C}} \left( p_c \cdot D_{av}^C(j) + p_d \cdot D_{av}^D(j) +
                                            p_s \cdot (D_{av}^D(j) + D_{max}^C(j') + D_{av}^C(j')) \right)$\\
   & $\leq \sum_{j \in \mathcal{C}} \left( (p_c + p_s) \cdot D_{av}^C(j) + (p_d + 2p_s) \cdot D_{av}^D(j) \right)$\\
   & $= \sum_{j \in \mathcal{C}} \left( (p_c + p_s) \cdot (C^*_j - r_\gamma'(j) \cdot F^*_j) +
                                                                            (p_d + 2p_s) \cdot (C^*_j + r_\gamma(j) \cdot F^*_j) \right)$\\
   & $= ((p_c + p_d + p_s) + 2p_s) \cdot C^* $\\
   & \hspace{3mm} $+ \sum_{j \in \mathcal{C}} \left( (p_c + p_s) \cdot ( - r_\gamma(j) \cdot (\gamma - 1) \cdot F^*_j)           + (p_d + 2p_s) \cdot ( r_\gamma(j) \cdot F^*_j) \right)$\\
   & $= (1 + 2p_s) \cdot C^* +
        \sum_{j \in \mathcal{C}} \left( F^*_j \cdot r_\gamma(j) \cdot (p_d + 2p_s - (\gamma - 1) \cdot (p_c + p_s))\right)$\\
   & $\leq (1 + \frac{2}{e^\gamma}) \cdot C^* +
        \sum_{j \in \mathcal{C}} \left( F^*_j \cdot r_\gamma(j) \cdot
        (\frac{1}{e} + \frac{1}{e^\gamma} - (\gamma - 1) \cdot (1 - \frac{1}{e} + \frac{1}{e^\gamma})) \right)$.\\
\end{tabular}
\end{center}

In the above calculation we used the following properties.
In the first inequality we explored the fact that cluster centers were chosen greedily,
which implies $D_{max}^C(j') + D_{av}^C(j') \leq D_{max}^C(j) + D_{av}^C(j)$.
For the last inequality, we used
$p_d + 2p_s = 1 - p_c + p_s \leq 1 - (1 - \frac{1}{e}) + \frac{1}{e^\gamma} = \frac{1}{e} + \frac{1}{e^\gamma}$.

It remains to observe that by setting $\gamma = \gamma_0 \approx 1.67736$ (see~(\ref{eq:gamma0})) we eliminate
the last term in the connection cost bound,
and we obtain $E[C_{SOL}] \leq (1 + \frac{2}{e^{\gamma_0}}) \cdot C^* \leq 1.37374 \cdot C^*$.
\qed
\end{proof}

\begin{figure}[t]

\hspace{-8mm}\includegraphics{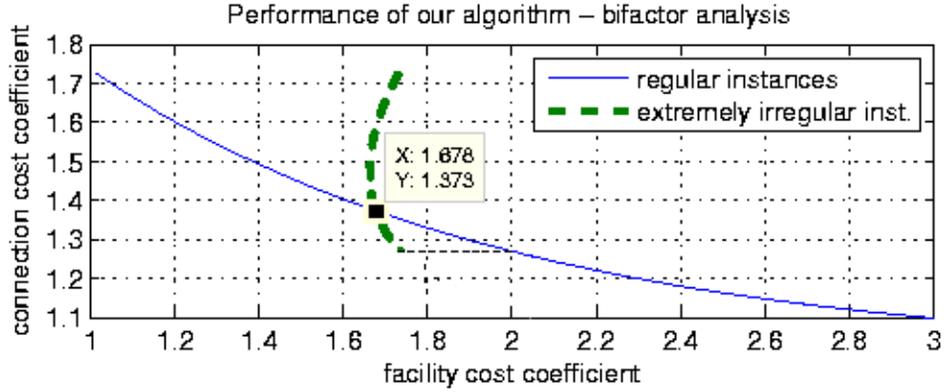}
\caption{The performance of our algorithm for different values of parameter $\gamma$.
The solid line corresponds to regular instances with $r_\gamma(j) = 0$ for all $j$ and it coincides with the
approximability lower bound curve. The dashed line corresponds to instances with $r_\gamma(j) = 1$ for all $j$.
For a particular choice of $\gamma$ we get a horizontal segment connecting those two curves;
for $\gamma \approx 1.67736$ the segment becomes a single point.
Observe that for instances dominated by connection cost only a regular instance may be tight for the lower bound.
}
\label{fig:performance}
\end{figure}

The algorithm A1($\gamma_0$) was described as a procedure of rounding a particular fractional solution
to the LP relaxation of the problem.
In the presented analysis we compared the cost of the obtained solution with
the cost of the starting fractional solution.
If we appropriately scale the cost function in the LP relaxation before solving the relaxation,
we easily obtain an algorithm with a bifactor approximation guaranty in a stronger sense. Namely,
we get a comparison of the produced solution with any feasible solution to the LP relaxation of the problem.
Such a stronger guarantee is, however, not necessary to construct the 1.5-approximation algorithm for
the metric UFL problem, which is presented in the next section.

The algorithm A1($\gamma$) with $\gamma = 1 + \epsilon$ (for a sufficiently small positive $\epsilon$)
is essentially the algorithm of Chudak and Shmoys. Observe that for regular instances, namely those with $r_\gamma(j) = 0$ for every client $j$, we do not need to set $\gamma = \gamma_0$ to eliminate the
dependence of connection cost of the produced solution on the facility opening cost of the fractional solution.
Hence, for regular instances, we get a $(\gamma, \frac{2}{e^\gamma})$-approximation algorithm
for each choice of $\gamma > 1$.

\section{The 1.5-approximation algorithm}\label{sec_approx}

In this section we will combine our algorithm with an earlier algorithm of Jain et al. to obtain
a 1.5-approximation algorithm for the metric UFL problem.

In 2002 Jain, Mahdian and Saberi~\cite{JMS} proposed a primal-dual approximation algorithm (the JMS algorithm).
Using a dual fitting approach they showed that it is a 1.61-approximation algorithm. Later
Mahdian, Ye and Zhang~\cite{mahdian-improved} derived the following result.

\begin{lemma}[\cite{mahdian-improved}] \label{MYZ_bifactor}
    The cost of a solution produced by the JMS algorithm is at most $1.11 \times F^* + 1.7764 \times C^*$,
    where $F^*$ and $C^*$ are facility and connection costs in an optimal solution to the linear relaxation of the problem.
\end{lemma}


\begin{theorem}
  Consider the solutions obtained with the A1($\gamma_0$) and JMS algorithms.
  The cheaper of them is expected to have a cost at most
  $1.5$ times the cost of the optimal fractional solution.
\end{theorem}

\begin{proof}
Consider an algorithm A2 that does the following. With probability $p=0.313$ runs the JMS algorithm
and otherwise, with probability $1-p$, runs the A1($\gamma_0$) algorithm.
Suppose that we are given an instance, and that $F^*$ and $C^*$ are facility and connection costs
in an optimal solution to the linear relaxation of this instance.
Consider the expected cost of the solution produced by algorithm A2 for this instance.
$
E[cost] \leq p \cdot (1.11 \cdot F^* + 1.7764 \cdot C^*) + (1-p) \cdot (1.67736 \cdot F^* + 1.37374 \cdot C^*)
    = 1.4998 \cdot F^* + 1.4998 \cdot C^* < 1.5 * (F^* + C^*) \leq 1.5 * OPT.
$
\qed
\end{proof}

Instead of the JMS algorithm we could take the algorithm of Mahdian et al.~\cite{mahdian-improved},
the MYZ($\delta$) algorithm, that scales the facility costs by $\delta$, runs the JMS algorithms, scales back the facility costs
and finally runs the greedy augmentation procedure.
With the notation introduced in Section~\ref{sub_greedy_aug},
the MYZ($\delta$) algorithm is the $S_\delta(JMS)$ algorithm.
The MYZ(1.504) algorithm was proven~\cite{mahdian-improved} to be
a 1.52-approximation algorithm for the metric UFL problem.
We may change the value of $\delta$ in the original analysis
to observe that MYZ(1.1) is a (1.2053,1.7058)-approximation algorithm.
This algorithm combined with our A1($\gamma_0$) (1.67736,1.37374)-approximation algorithm
gives a 1.4991-approximation algorithm for UFL.
This shows how much improvement we obtain by using the scaling
technique on the greedy algorithm's side.

\section{Multilevel facility location} \label{sec_multilevel}

In the $k$-level facility location problem the clients need to be connected to open facilities
on the first level, and each open facility except on the last, $k$-th level, needs to be connected
to an open facility on the next level. Aardal, Chudak, and Shmoys~\cite{Aardal99} gave a 3-approximation
algorithm for the $k$-level problem with arbitrary $k$. Ageev, Ye, and Zhang~\cite{Ageev03} proposed a reduction
of a $k$-level problem to a $(k-1)$-level and a $1$-level problem, which results in a recursive algorithm. This algorithm
uses an approximation algorithm for the single level problem and
has a better approximation ratio, but only for instances with small $k$. Using our new
algorithm A1($\gamma_0$) instead of the JMS algorithm within this framework,
improves approximation for each level. In particular, in the limit as $k$ tends to $\infty$,
 we get a 3.236-approximation which is the best possible for this construction.

By a slightly different method, Zhang~\cite{Zhang06} obtained a 1.77-approximation algorithm
for the 2-level problem. For the 3-level and the 4-level version of the problem
he obtained 2.523-\footnote{This value deviates slightly
from the value 2.51 given in the paper. The original argument contained a minor calculation error.}
and 2.81-approximation algorithms, by reducing to a problem with smaller number of levels.
In the following section we will modify the algorithm by Zhang for the 3-level problem, and use the new
(1.67736,1.37374)-approximation algorithm for the single-level part, to obtain a 2.492-approximation,
which improves on the previously best known approximation by Zhang.
Note, that for $k>4$ the best known approximation factor is still due to Aardal et al.~\cite{Aardal99}.

\subsection{3-level facility location}
We will now present the ingredients of the 2.492-approximation algorithm.
We start from an algorithm to solve the 2-level version.

\begin{lemma}[Theorem 2 in~\cite{Zhang06}] \label{lem:3l1}
The 2-level UFL problem may be approximated by a factor of $1.77+\epsilon$ in polynomial time
for any given constant $\epsilon > 0$.
\end{lemma}

Zhang~\cite{Zhang06} also considered a scaling technique
analogous to the one described in Section~\ref{sub_greedy_aug},
but applicable to the 2-level version of the problem.
An effect of using this technique is analyzed in the following
lemma.

\begin{lemma}[Theorem 3 in~\cite{Zhang06}] \label{lem:3l2}
For any given $\epsilon > 0$, if there is an $(a,b)$-approximation algorithm
for the 2-level UFL problem, then we can get an approximation algorithm
for the 2-level UFL problem with performance guarantee
\[
 \left( a + \frac{e}{e-1}ln(\Delta) + \epsilon, 1 + \frac{b-1}{\Delta} \right)
\]
for any $\Delta \geq 1$.
\end{lemma}

He also uses the following reduction.

\begin{lemma}[Lemma 7 in~\cite{Zhang06}] \label{lem:3l3}
Assume, that the 1-level and 2-level UFL problems
have approximation algorithms with factors
$(a,b)$ and $(\alpha, \beta)$, respectively, then the 3-level
UFL problem may be approximated by factors $(max\{a, \frac{a+\alpha}{2}\}, \frac{3b + \beta}{2})$.
\end{lemma}

Zhang~\cite{Zhang06} observed that the above three statements
may be combined with the MYZ algorithm to improve
the approximation ratio for the 3-level UFL problem.
In the following theorem we show that we may
use our new (1.6774,1.3738)-approximation algorithm for the
1-level UFL problem to get even better approximation
for the 3-level variant.

\begin{theorem}
 There is a 2.492-approximation algorithm
 for the 3-level UFL problem.
\end{theorem}
\begin{proof}
We first use the algorithm from Lemma~\ref{lem:3l1},
and the scaling technique from Lemma~\ref{lem:3l2}, with $\Delta = 1.57971$,
to obtain a $(2.492, 1.48743)$-approximation algorithm for the
2-level UFL problem.

Then we use our $(1.6774,1.3737\ldots)$-approximation algorithm for the
1-level UFL problem with the scaling technique from Lemma~\ref{lem:scaling},
with $\gamma = 2.25827$,
to obtain a $(2.492,1.1655)$-approximation algorithm for the 1-level UFL problem.

Finally, we use Lemma~\ref{lem:3l3} to combine these two algorithms
into a $(2.492, 2.492)$-approximation algorithm for the
3-level UFL problem.
\qed
\end{proof}

\section{Universal randomized clustering procedure}\label{sec_randomized}

In this section we discuss a different approach to clustering.
We propose to modify the greedy clustering algorithm by choosing consecutive cluster centers
randomly with uniform distribution. The output of such a
process is obviously random, but we may still prove some statements about probabilities.
A resulting clustering will be denoted by a function $g : \mathcal{C} \to \mathcal{C}$,
which assigns to each client $j$ the center of its cluster $j' = g(j)$.
The following lemma states that the clustering $g$ obtained with the
randomized clustering procedure is expected to be ``fair''.

\begin{lemma} \label{random_clustering_lemma}
  Given a graph $G=(\mathcal{F} \cup \mathcal{C}, E)$ and assuming that a clustering $g$ was obtained
  by the above described random process, for every two distinct clients $j$ and $j'$,
  the probability that $g(j)=j'$ is equal the probability that $g(j')=j$.
\end{lemma}

\begin{proof}
Let $C(G)$ denote the maximal (over the possible random choices of the algorithm)
number of clusters that can be obtained from $G$ with the random clustering procedure.
The proof will be by induction on $C(G)$. Fix any $j,j'\in \mathcal{C}$ such that $j$ is a neighbor of $j'$ in $G$
(if they are not neighbors, neither $g(j)=j'$ nor $g(j')=j$ can occur).
Suppose $C(G) = 1$, then $Pr[g(j)=j'] = Pr[g(j')=j] = 1/|\mathcal{C}|$.

Let us now assume that $C(G) > 1$. There are two possibilities,
either one of $j,j'$ will belong to the first cluster, or none of them will.
Consider the first case (the first chosen cluster center is either $j$ or $j'$ or one of their neighbors).
If $j$ ($j'$) is chosen as a cluster center, then $g(j')=j$ ($g(j)=j'$).
Since they are chosen with the same probability, the contribution of the first case
to the probability of $g(j')=j$ is equal to the contribution to the probability of $g(j)=j'$.
If neither of them gets chosen as a cluster center but at least one belongs to the new cluster,
then neither $g(j')=j$ nor $g(j)=j'$ is possible.

Now consider the second case (neither $j$ nor $j'$ belongs to the first cluster).
Consider the graph $G'$ obtained from $G$ by removing the first cluster.
The random clustering proceeds like it has just started with the graph $G'$,
but the maximal number of possible clusters is smaller: $C(G') \leq C(G) -1$.
Therefore, by the inductive hypothesis, in a random clustering of
$G'$ the probability that $g(j')=j$ is equal to the probability that $g(j)=j'$.
%
%
%
\qed
\end{proof}

If $g(j)=j'$ in a clustering $g$ of graph $G$ we will say that client $j'$ \emph{offers support} to client $j$.
The main idea behind the clustering algorithms for the UFL problem is that we may afford
to serve each cluster center directly (because they are never neighbors in $G$) and all the
other clients are offered support from their cluster centers. A non-central client may either
accept the support and connect itself via its cluster center (that is what all non-central clients
do in the algorithm of Shmoys et al.), or it may try to get served locally, and if it fails,
accept the support (this is the way the Chudak and Shmoys' algorithm works). In both those algorithms
the probability that an offer of support is accepted is estimated to be constant.
Therefore, we may modify those algorithms to use the random clustering procedure
and do the following analysis.

For any two clients $j$ and $j'$, the probability
that $j$ accepts the support of $j'$ is equal to the probability that $j'$ accepts the
support of $j$. Let $i$ be a facility on a shortest path from $j$ to $j'$.
When we compute the expected connection cost of client $j$, we observe that with certain probability $p$
it accepts the support of $j'$. In such a case it must pay for the route via $i$ and $j'$ to the facility
directly serving $j'$. We will now change the bookkeeping and say that in this situation $j$ is paying only for the part until facility $i$, and the rest is paid by $j'$,
but if $j$ would be supporting $j'$ it would have to pay a part of $j'$'s connection cost,
which is the length of the path from $i$ via $j$ to the facility serving $j$.
We may think of this as each client having a bank account, and when it accepts support
it makes a deposit, and when it offers support and the support is accepted, then
it withdraws money to pay a part of the connection cost of the supported client.
From Lemma~\ref{random_clustering_lemma} we know that for a client $j$
the probability that it will earn on $j'$ is equal to the probability that
it will lose on $j'$. Therefore, if the deposited amount is equal to the withdrawal,
the expected net cash flow is zero.

The above analysis shows that randomizing the clustering phase of the known LP-rounding algorithms would
not worsen their approximation ratios. Although it does not make much sense to use a randomized algorithm
if it has no better performance guarantee, the random clustering has an advantage of allowing the analysis
to be more local and uniform.

\section{Concluding remarks}\label{sec_concluding}

With the 1.52-approximation algorithm of Mahdian et al. it was not clear to the authors
if a better analysis of the algorithm could close the gap with the approximation lower bound of 1.463
by Guha and Khuller. In~\cite{Byrka06} we have recently given a negative answer to this question
by constructing instances that are hard for the MYZ algorithm. Similarly, we now do not know if our new
algorithm $A1(\gamma)$ could be analyzed better to close the gap. Construction of hard instances for our algorithm
remains an open problem.

The technique described in Section~\ref{sub_greedy_aug} enables us
to move the bifactor approximation guarantee of an
algorithm along the approximability lower bound of Jain et al.
(see Figure~\ref{fig_contr}) towards higher facility opening costs.
If we developed a technique to move the analysis in the opposite direction,
together with our new algorithm, it would imply
closing the approximability gap for the metric UFL problem.
It seems that with such an approach we would have to face the
difficulty of analyzing an algorithm that closes some of the previously opened facilities.

\section*{Acknowledgments}
The authors want to thank David Shmoys, Steven Kelk, Evangelos Markakis
and the anonymous referees of this submission as well as the referees
of the earlier conference version of this paper \cite{Byrka07}
for their advice and valuable remarks.

\bibliographystyle{abbrv}
\bibliography{}

\end{document}